# INFANT CRY CLASSIFICATION WITH GRAPH CONVOLUTIONAL NETWORKS

*Chunyan Ji[1], Ming Chen[2], Bin Li[3], Yi Pan[1]*

[1]Department of Computer Science, Georgia State University, Atlanta, USA
[2]College of Information Science and Engineering, Hunan Normal University, Changsha, China
[3]School of Computer Science and Technology, Anhui University, Hefei, China

## ABSTRACT

We propose an approach of graph convolutional networks for robust infant cry classification. We construct non-fully connected graphs based on the similarities among the relevant nodes in both supervised and semi-supervised node classification with convolutional neural networks to consider the short-term and long-term effects of infant cry signals related to inner-class and inter-class messages. The approach captures the diversity of variations within infant cries, especially for limited training samples. The effectiveness of this approach is evaluated on Baby Chillanto Database and Baby2020 database. With as limited as 20% of labeled training data, our model outperforms that of CNN model with 80% labeled training data and the accuracy stably improves as the number of labeled training samples increases. The best results give significant improvements of 7.36% and 3.59% compared with the results of the CNN models on Baby Chillanto database and Baby2020 database respectively.

## 1. INTRODUCTION

Infants communicate with the world by crying, which contains many reasons such as pain, discomfort, and hunger, etc. Previous work shows that infant crying is a short-term stationary signal and only contains non-speech information [1]. Infant crying is a combination of vocalization, silence, coughing, choking, and interruptions, which includes a diversity of acoustic and prosodic information at different levels. The infant crying falls in the most sensitive range of the human auditory sensation [2].

Neural Networks (NNs) have been applied for infant cry classification and feature extraction. The bottleneck approach extracts tandem features through a deep NN, and then stitch it together with original Mel Frequency Cepstral Coefficient (MFCC) features to further improve the recognition accuracy [3]. The pretrained Convolutional Neural Networks (CNNs) as feature extractors on out-of-domain data set has a good performance [4]. Methods from image processing are applied to infant cry classification achieving good performances [5][6]. Transfer learning approaches are effective in transferring rich phonetic and acoustic information from adults' model to children model [7]. Features in different regions have different transferability in CNN acoustic model across languages [8]. The use of Transfer Learning (TL) CNN to extract the palmprint image features and feed into the SVM classifier was beneficial for limited data [9]. TL CNN with spectrograms and SVM model fusion method is used to improve the infant cry reason classification performance over the traditional CNN model [10]. The cutting-edge method of Graph Neural Network (GNNs) demonstrated ground-breaking performance on many tasks. The connectionist models capture the dependence of graphs via message passing between the nodes of graphs [11]. Graph Convolutional Networks (GCNs) is a branch of GNNs that has a good balance of spectral and spatial representation [12]. To date, GNN is rarely used in audio research domain yet and a thorough search yields two relevant studies. GCN with a fully connected graph was used to classify music genre and achieves high accuracy [13]. The attentional GNN based few-shot learning was proposed in environmental sound classification achieving satisfying improvements [14]. Previous NN based methods address the fundamental work of infant cry classification. Challenges remain in this domain, especially under the limited labeled data with inconsistent transcriptions.

The available labeled infant cry data is very limited because data acquisition is sensitive and transcribing such raw data is time consuming and requires pediatricians or experienced care givers. Moreover, infant cry is more stationary and includes complicated acoustic and prosodic phenomena related to short-term and long-term influence. The uncertainty and inconsistency of the labels are higher than that of speech because the meaning behind infant cries are complex and possibly mixed. Recently, CNN approaches improve the performance for infant cry classification, whereas the use of fixed filter sizes in CNN cannot consider the long-term effects within the stationary infant cry signals. Transfer learning is powerful on image processing, whereas spectrograms are quite different from images. It is essential to discover novel efficient approaches to learn the short-term and long-term effects among cries for robust infant cry classification under limited samples recorded in real environments.

In this paper, we propose applying GCN for infant cry classification. We establish the graphs connected by weighted edges based on the similarities of the relevant nodes. The non-fully connected graphs feeding to the GCN is constructed taking into account the short-term and long-term effects of infant cry signals related to inner-class and inter-class messages. Our GCN approach captures the long-distance variations within infant cries, especially under the limited training samples. The effectiveness of proposed method is evaluated on commonly used Baby Chillanto database and a newly collected Baby2020 database. Our major contributions include: (1) we apply GCN on robust infant cry classification and achieve significant improvements; (2) our approach with 20% of labeled data outperforms the CNN model with 80% of labeled data; (3) the novel non-fully connected graph with GCN is efficient to capture the long-distance effects in infant cry for better discriminative ability; (4) the challenge of limited training data with uncertain labels can be solved using weighted class-crossing edges in the graph of GCNs.

The remainder of the paper is organized as follows. In Section 2, we compare the differences between infant cries and speech signals both in time and frequency domain. Section 3 outlines our proposed approach of infant cry classification with GCN. Section 4 describes our experimental results and we conclude in Section 5.

## 2. INFANT CRY VS. SPEECH

The main components of a cry sound are expiration and inspiration segments with vocalization and audible expiration and inspiration [15]. The dominant frequency components of infant cry mainly range from 1500 Hz to 3000 Hz while it is mainly distributed below 800 Hz for male adult speech and the power spectrum of infant cry is more continuous compared with that of adult voices [16]. Figure 1 demonstrates a comparison of waveforms and spectrograms of an infant cry and a male adult speech. Infant cry has higher fundamental frequency F0 within 250-700Hz compared with that of adult speech ranged 85Hz to 200Hz. Figure 1 (a) shows that envelop of the intensity of infant cry signal is rhythmic and contains cyclic changes due to the natural breath. Infant cry has high pitch and has periodic nature alternating crying and respirations. The pitch contour as well as the envelops of F123 formants from infant cry signals are relatively flat compared to speech signals. It means the tone of infant crying is consistent and the fluctuations are relatively shorter. Figure 1 shows the spectrograms are images including both acoustic and prosodic information. The prosodic features include intensity, F0, formants, etc., which are essential to describe the variations in the infant cry signals.

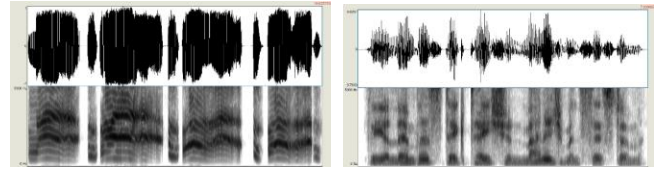

a: Waveform and spectrogram of an infant cry   b: Waveform and spectrogram of a male adult speech

Fig.1: Infant cry vs. adult speech in time and frequency domain.

## 3. INFANT CRY CLASSIFICATION WITH GCN

### 3.1. The structure of proposed approach

The architecture of our model for infant cry classification is shown in Figure 2. It includes three parts: spectrogram generation, TL CNN feature extractor, and GCN classification. As shown in Figure 2, spectrograms are generated from the audio wav samples and are fed into the TL CNN feature extractor to extract multi-dimensional feature vectors, which are the input of the GCN classifier. Both supervised and semi-supervised audio classification with GCN are performed in the experiments.

Since spectrograms have strong relationship with images and ResNet50 [17] is proven to perform well on image classification, we feed the spectrograms into the feature extractor, which is built using a ResNet50 based TL CNN model. The layers before the fully connected layer in ResNet50 are saved as the base model of the feature extractor. Appending some custom layers after the base model, we train the model using the training set and then extract the feature vectors of the testing set from the last fully connected layer when predicting the test set. The training and extracting are performed multiple times using different data to extract the features of the whole dataset.

### 3.2. Graph Convolutional Network node classification

As CNN being widely used for images classification, it learns the features within the images. By passing messages and aggregating features among neighbors, GCN is a promising choice to learn hidden relationship among images. GCN node classification takes a graph as input and determine labels of nodes by learning the features of their own and associated neighbors. The input graph is defined as $G = (V, \mathbf{A})$, where $V$ represents the vertex set consisting of nodes $\{v_1, \ldots, v_n\}$ and $\mathbf{A} \in \mathrm{R}^{n \times n}$ is the adjacency matrix where $a_{ij} \geq 0$ denotes the edge weight between nodes $v_i$ and $v_j$. The GCN we consider in our experiments have the similarity of each pair of images as the edge weight and each node $v_i$ has a corresponding *1024*-dimensional feature vector $x_i \in R^{1024}$. Each entry on the diagonal degree matrix $D = Diag(d_i, \ldots, d_n)$ is equal to the row-sum of the adjacency matrix $d_i = \sum_0^j a_{ij}$.

GCN learns a new feature representation for the feature $x_i$ of each node over multiple layers and subsequently use it as input into a linear classifier. In each graph convolution layer, the features $x_i$ of each node $v_i$ are updated via averaging within its local neighborhood by feature propagation:

$$x_i^{(k)} \leftarrow \frac{1}{d_i+1} x_i^{(k-1)} + \sum_{j=1}^{n} \frac{1}{\sqrt{d_i+1}\sqrt{d_j+1}} x_i^{(k-1)} \quad (1)$$

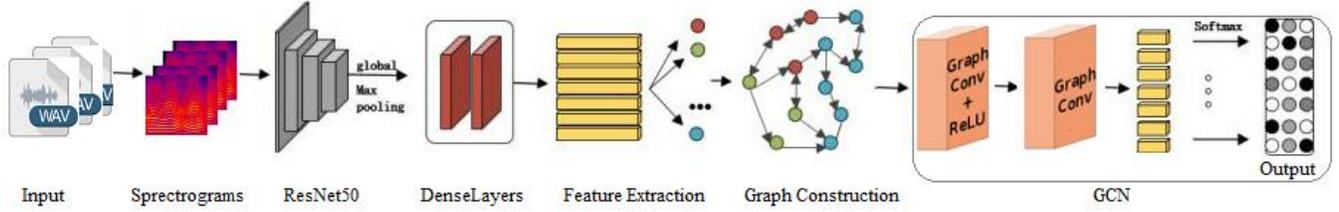

Fig. 2: Our proposed model for infant cry classification with GCN.

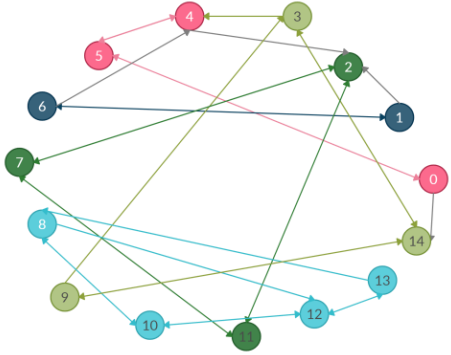

Fig. 3: Graph constructed for GCN.

Intuitively, this step smooths the hidden representations locally along the edges of the graph and ultimately encourages similar predictions among locally connected nodes. After the local smoothing, the operation goes to a standard Multi-Layer Perceptron (MLP) with a weight matrix $\Theta^{(k)}$ and a nonlinear activation function, such as ReLU in our experiment, is applied pointwise before outputting the new feature representation $\mathbf{X}^{(k)}$. The last layer of a GCN predicts the labels using the softmax function, by which each node belongs to one out of $C$ classes and can be labeled with a $C$-dimensional one-hot vector $y_i \in \{0, 1\}^C$.

### 3.3. Graph construction for GCN

The feature matrix extracted from the TL CNN is then used to calculate the similarity of each pair of nodes to construct the graph for GCN in semi-supervised and supervised methods.

- Semi-supervised graph construction for GCN. The graph fed to the GCN is built based on the similarities calculated using all samples including labeled training set and unlabeled testing set. We calculate the similarity of each pair of nodes by their Euclidean distance. For each node, the similarities to all other nodes are ordered and a hyperparameter $k$ is tuned to decide the degree of the edges. Once two nodes are connected, the similarity between them is used as the weight on the corresponding edge. The weight matrix shows the similarity values if two nodes are connected and 0 value otherwise.

- Supervised graph construction for GCN. Different from semi-supervised method, supervised method builds the training graph and the testing graph separately based on the training set and testing set ratio. For example, 80% data is used to construct training graph and 20% is used to construct testing graph. Same as semi-supervised method, the graph is constructed by connecting $k$ number of closest neighbors from each node. After training, the testing graph is fed into the trained GCN model for node classification.

Figure 3 describes a graph constructed before feeding to the GCN. It has 15 nodes in five classes with $k$ value equals to 2 and each node contains a feature vector extracted from the TL CNN. Double arrowed edges represent both nodes are the $k$ closest neighbors of each other. Single arrow means only the arrow-side node is the $k$ closest neighbor of the node without the arrow, but not the other way around. For example, node 0 and node 5 are close to each other while node 2 is close to node 1, but node 1 is not close to node 2. In the experiments, each edge is associated with a weight, which is the similarity calculated by the Euclidean distance between the feature vectors of the two nodes.

GCN with supervised method and semi-supervised method are both valuable in infant cry classification tasks due to the lack of human labeled samples. Discovering effective semi-supervised method can help make good use of large amount of unlabeled cry samples. Semi-supervised learning is important as it can leverage ample available unlabeled data to aid supervised learning, thus greatly saving the cost, trouble, and time for human labeling. Graph-based methods have been demonstrated as one of the most effective approaches for semi-supervised learning, as they can exploit the connectivity patterns between labeled and unlabeled data samples to improve learning performance. In GCN setting, semi-supervised classification has advantages over supervised classification because semi-supervise graph is constructed by the whole dataset, which means the distances among the testing data and training data are also calculated. Connecting unlabeled testing nodes to labeled training nodes before training brings some level of prior knowledge to the unlabeled data samples.

## 4. EXPERIMENTAL RESULTS AND ANALYSIS

### 4.1. Datasets and experimental setup

We evaluate our approach on Baby Chillanto database and Baby2020 dataset with larger size. Baby Chillanto database is collected by the National Institute of Astrophysics and Optical Electronics, CONACYT, Mexico [18]. The babies being recorded are ranging from newborn to nine months of age and each sample is a one-second long audio wav file. The total number of cries we use in this work is 2267 including 340 asphyxia, 879 deaf, 350 hunger, 507 normal, and 191 pain labeled by medical doctors. Baby2020 database is our developing database with 50 hours of recordings collected from over 100 babies. We select 5540 cry samples of three types (2880 hungry, 1700 Sleepy, 960 Wakeup) from healthy infants of 0 to 3 months old. The cries are recorded via mobile devices placed right beside the infants in real environments and labeled either by parents at home or by doctors and nurses in hospitals. Cry samples are manually segmented with the length between 1 second to 7 seconds.

Sound eXchange (Sox) software [19] is used to generate spectrograms from the wav file samples. Each spectrogram image is in size 256×256. From the trained ResNet50, we take the convolutional layers before the fully connect 1000 layer as the base model. Then, we append custom layers, a GlobalMaxPooling layer, two dense layers with 1024 neurons, a dropout layer with a rate of 0.25, and another 1024-neuron dense layer, to the base ResNet50 model. The TL CNN model is trained on 80% samples and used to extract the features for the rest of the 20% testing samples. The total 2267 samples are divided into 5-fold training and testing sets. After five times training and extracting, the features for all 2267 samples are extracted, and then are combined to be the feature vectors stored in a csv file, which is a 2267×1024 matrix. For Baby2020 dataset, we get a 5540×1024 feature matrix.

The feature matrix is then used to calculate similarity of each pair of nodes and the graphs are constructed for GCN, which is introduced in [12]. The hyperparameters are tuned and the same values are used for all experiments on the same dataset. The values of the hyperparameters on Baby Chillanto database are: 2-layer GCN, 32 number of hidden neurons in GCN, 2000 epochs, 0.001 learning rate, 0.1 dropout, and $k$ value is 3. The hyperparameter values for Baby2020 database are the same except the $k$ value is equal to 1 to produce the best performance. All experiments are performed with 5-fold cross validation to ensure the reliability of the accuracy.

### 4.2. Experimental results

Table 1 shows our method outperforms other methods on Baby Chillanto database using spectrograms as input. Comparing to the standard CNN model, our approach in supervised and semi-supervised settings improves testing accuracy by 4.98% and 7.36% respectively. Our GCN approach aggregates similar features with short and long distance in the time series. Aggregating features of nodes in different classes can also help soothe the mislabeling effect, which is unavoidable because cry reasons may be mixed in one cry sample. Semi-supervised graph provides some prior knowledge of the relationship between testing data and the labeled training data, hence, produces better accuracy than the supervised classification with GCN.

| Features | Model | Accuracy | Improvement |
|---|---|---|---|
| Spectrogram | CNN [10] | 87.03% | baseline |
| Spectrogram | TLCNN & SVM [10] | 90.80% | +3.77% |
| Spectrogram → TL CNN | GCN (supervised) | 92.01% | +4.98% |
| Spectrogram → TL CNN | GCN (semi-supervised) | 94.39% | +7.36% |

Table 1: Results comparison on Baby Chillanto database

| Features | Model | Accuracy | Improvement |
|---|---|---|---|
| Spectrogram | CNN | 70.78% | baseline |
| Spectrogram → TL CNN | GCN (supervised) | 74.24% | +3.46% |
| Spectrogram → TL CNN | GCN (semi-supervised) | 74.37% | +3.59% |

Table 2: Classification results on Baby2020 database

Table 2 shows the performances of our approach on Baby2020 dataset. It outperforms standard CNN model in both supervised and semi-supervised settings by improving accuracy 3.46% and 3.59% respectively. The classification accuracy on Baby2020 dataset is not as high as the one on Baby Chillanto database. We believe it is because data samples recorded by recording software installed on mobile devices are in mp3 and m4a formats, which use lossy compression technique causing loss of some high frequency information of the audios. The recording environments also have some background noises and the three types of crying, hungry, sleepy, and wakeup, are from healthy infants and have high level of similarity. The effectiveness of GCN on such dataset indicates that the relationship among the same class and different class data samples are considered in learning and help improve the learning outcomes. The Baby2020 dataset will be our focus in the future because it has the same setting as real time infant cry classification system, by which the user can catch infant cry signal using mobile devices and immediately tell infants' needs.

Table 3 and Figure 4 show that even with as limited as 20% of labeled data, our model outperforms that of the CNN model with 80% labeled training data and the accuracy stably improves as the number of labeled training samples increases in both semi-supervised and supervised settings. The semi-supervised method performs better because the graph contains more knowledge among labeled data and unlabeled testing samples.

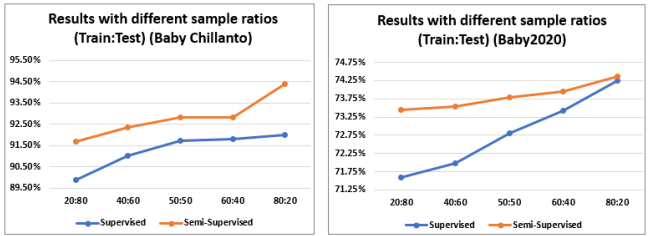

Fig. 4: Results of experiments with different sample ratios.

| Dataset | Model (Train:Test) | Accuracy | Improvement |
|---|---|---|---|
| **Chillanto** | ***CNN (80:20)*** | ***87.03%*** | --- |
| Chillanto | GCN (supervised 20:80) | 89.89% | +2.86% |
| Chillanto | GCN (supervised 50:50) | 91.73% | +4.70% |
| Chillanto | GCN (supervised 80:20) | 92.01% | +4.98% |
| Chillanto | GCN (semi-supervised 20:80) | 91.68% | +4.65% |
| Chillanto | GCN (semi-supervised 50:50) | 92.80% | +5.77% |
| Chillanto | GCN (semi-supervised 80:20) | **94.39%** | **+7.36%** |
| ***Baby2020*** | ***CNN (80:20)*** | ***70.78%*** | --- |
| Baby2020 | GCN (supervised 20:80) | 71.58% | +0.8% |
| Baby2020 | GCN (supervised 50:50) | 72.81% | +2.03% |
| Baby2020 | GCN (supervised 80:20) | 74.24% | +3.46% |
| Baby2020 | GCN (semi-supervised 20:80) | 73.45% | +2.67% |
| Baby2020 | GCN (semi-supervised 50:50) | 73.79% | +2.75% |
| Baby2020 | GCN (semi-supervised 80:20) | **74.37%** | **+3.59%** |

Table 3: Main results of experiments with different sample ratios

## 5. CONCLUSION

We demonstrated that using GCN with transfer learning CNN extracted features measured with similarity is efficient for robust infant cry classification. The non-fully connected graphs of GCN are constructed to consider the short-term and long-term effects of infant cry signals related to inner-class and inter-class messages. The improved discriminative ability of local nodes has the benefits of capturing the long-distance variations within infant cries, especially under the limited training data. The experimental results show that even with as limited as 20% of labeled data, our model outperforms that of the CNN model with 80% labeled training data and the accuracy increases as more training samples are used. The best accuracy improves 7.36% and 3.59% on Baby Chillanto database and Baby2020 database respectively. Our future work includes generating reliable graph edges for signed GCNs via the use of prior knowledge of class, investigating the effects of different similarity kernels for edges, as well as designing a deeper network architecture for more infant cry samples. Our approach outperforms other methods in infant cry classification and can be easily extended to other acoustic event classification tasks.

## ACKNOWLEDGMENT

We want to express our great gratitude to Dr. Orion Reyes and Dr. Carlos A. Reyes for providing the access to the Baby Chillanto database. We want to thank all parents, doctors, and nurses who support the recording of Baby2020 database. We also want to gratefully acknowledge the support of NVIDIA Corporation with the donation of the Tesla K40 GPU used for this research.